\documentclass[aps,prb,twocolumn,superscriptaddress,reprint]{revtex4-2}

\usepackage{float}
\usepackage{graphicx}
\usepackage{amsmath}
\usepackage{bm}
\usepackage{cases}
\usepackage{color}
\usepackage{amssymb}
\usepackage{mathrsfs}
\usepackage{dcolumn}
\usepackage{multirow}
\usepackage{wasysym}
\usepackage{tabularx} 
\usepackage[mathlines]{lineno}
\usepackage[colorlinks=true,linkcolor=blue,urlcolor=blue,citecolor=blue]{hyperref}

\begin{document}


\title{Klein Tunneling of Dirac Fermions through Electromagnetic Barriers}

\author{Lingang Zhang}
\affiliation{Department of Physics, Zhejiang Normal University, Jinhua 321004, China}

\author{Hua Chen}
\email{Electronic address: hwachanphy@zjnu.edu.cn} 
\affiliation{Department of Physics, Zhejiang Normal University, Jinhua 321004, China}

\begin{abstract}		
The Lorentz covariance of relativistic Dirac equations serves as a fundamental principle underlying the laws of electromagnetism across different inertial frames.
Exploiting the covariance, we obtain the general solutions for Dirac fermions under both the in-plane electric $\boldsymbol{E}$ and perpendicular magnetic $\boldsymbol{B}$ fields, which reduce to either a magnetic or electric field in the inertial frame with drift velocity along the $\boldsymbol{E}\times\boldsymbol{B}$ direction. This dichotomy defines the magnetic and electric regimes, separated by the critical field ratio $E/B=v_\text{F}$ with $v_\text{F}$ denoting the Fermi velocity of Dirac fermions.
Using these solutions, we revisit Klein tunneling through a heterojunction with generalized electromagnetic potentials.
In the magnetic regime, the transmission exhibits oscillations governed by the Fabry-P\'erot interference. In the electric regime, perfect transmission occurs at normal incidence in the drifted frame.
The interference phase is further analyzed in terms of the solid angles on the Bloch sphere, providing a geometric interpretation of Klein tunneling.
Finally, we briefly discuss the relation between the tilting of Dirac cones and the in-plane electric field, establishing the correspondence of the undertilted and overtilted cases to the magnetic and electric regimes, respectively.
Our findings reveal the manipulation of Klein tunneling by electromagnetic fields, offering a theoretical basis for designing novel electronic devices.	
\end{abstract}

\date{\today}

\maketitle


The linearly dispersed Dirac fermions, originally predicted as  relativistic particles~\cite{Dirac1928,Dirac1928a}, have ignited an explosion of research interest following their experimental discovery in solid-state materials~\cite{Novoselov2004,Berger2004,Novoselov2005,Zhang2005,CastroNeto2009,Peres2010,DasSarma2011,Goerbig2011,Vafek2014,Wehling2014,Lv2021}. 
In particular, a striking consequence of this relativistic nature is manifested in the phenomenon of Klein tunneling~\cite{Klein1929,Dombey1999,Allain2011}. In stark contrast to the non-relativistic paradigm, wherein an electron incident upon an electrostatic barrier experiences exponentially suppressed transmission, Dirac fermions exhibit perfect transmission through barriers of arbitrary height and width at normal incidence~\cite{Katsnelson2006a,Beenakker2008,Young2009,Stander2009,Young2011}. This anomalous transparency, rooted in the conservation of chirality and the concomitant prohibition of backscattering~\cite{Ando1998}, unequivocally distinguishes relativistic particles from their non-relativistic counterparts. 
Theoretically, the relativistic nature originates from the Lorentz covariance of Dirac equations~\cite{Dirac1928,Dirac1928a}, a fundamental principle that dictates the transformation of electromagnetic fields across different inertial frames~\cite{Jackson2009}.
Generally, the classical motion of Dirac fermions under both in-plane electric $\boldsymbol{E}$ and perpendicular magnetic $\boldsymbol{B}$ fields can be characterized by the open and closed trajectories in the electric and magnetic regimes, respectively~\cite{Shytov2009}. These two distinct regimes are demarcated by the critical ratio of electromagnetic fields $E/B=v_\text{F}$, where $v_\text{F}$ denotes the Fermi velocity of Dirac fermions.
Quantum mechanically, the Landau level spacing gradually contracts and eventually collapses as the critical field ratio $E/B=v_\text{F}$ is approached from the magnetic regime~\cite{Lukose2007}. In contrast, the corresponding behavior in the electric regime remains largely unexplored.

Here we derive the general solutions of arbitrary electromagnetic fields by exploiting the Lorentz covariance of Dirac equations, and revisit Klein tunneling across a heterojunction, in which the sharp electrostatic barriers are reshaped as generic electromagnetic potentials.
The condition for perfect Klein tunneling at normal incidence is  satisfied at the critical field ratio $E/B=v_\text{F}$, which corresponds to the simultaneous cancellation of electromagnetic fields in the drifted inertial frame.
In the magnetic regime, the oscillations of transmission, which are expected from the interference of Aharonov-Bohm phase, are characterized by the geometric phase of spinor wave functions at the interfaces of heterojunction, as visualized on the Bloch sphere.
In the electric regime, perfect Klein tunneling deviates from normal incidence to the angle at which the Lorentz force is balanced by the projected component of the electric force.
Finally, we demonstrate that the tilting of Dirac cones~\cite{Kino2006,Goerbig2008,Morinari2009} can be reformulated as an effective electric field~\cite{Farajollahpour2019}, thereby extending the applicability of our results to a broader range of materials.


\begin{table*}[]
	\caption{\label{tab} 
		General solutions of massless relativistic Dirac equation in Eq.~(\ref{eq:diraceq}) under both in-plane electric and the perpendicular magnetic fields. 
		Two linearly independent solutions in the magnetic and electric regimes are expressed in terms of the parabolic cylinder function $D_{\nu}$ and the Kummer function $M$, respectively~\cite{NIST:DLMF}.
		}
	\begin{ruledtabular}
		\begin{tabular}{lcc}
			\parbox{2.5cm}{}
			& Magnetic Regime ($E/B<v_\text{F}$)
			& Electric Regime ($E/B>v_\text{F}$)\\ \hline
			
			\parbox{2.5cm}{\raggedright Lorentz Transformation}
			& $\tanh\theta=\beta=E/v_\text{F}B$  
			& $\tanh\theta=\beta=v_\text{F}B/E$ \\ \hline
			
			\parbox{2.5cm}{\raggedright Effective Fields in Drifted Frame}
			& $\begin{aligned}
				\boldsymbol{E^{\prime}}&=0\\
				\boldsymbol{B^{\prime}}&=B\sqrt{1-\beta^{2}}\hat{z}
			\end{aligned}$ 
			& $\begin{aligned}
				\boldsymbol{E^{\prime}}&=E\sqrt{1-\beta^{2}}\hat{x}\\
				\boldsymbol{B^{\prime}}&=0
			\end{aligned}$ \\ \hline
			
			\parbox{2.5cm}{\raggedright Characteristic Length}
			& $l_{B}^{\prime}=\sqrt{\hbar/eB^{\prime}}$  
			& $l_{E}^{\prime}=\sqrt{\hbar v_{F}/eE^{\prime}}$ \\ \hline
			

			\parbox{2.5cm}{\raggedright Principal Quantum Number}
			&$\nu=\cfrac{\epsilon^{\prime2}l^{\prime2}_{B}}{2\hbar^{2}v^{2}_{\text{F}}}-1$
			& $\epsilon^\prime$ \\ \hline

			\parbox{2.5cm}{\raggedright Wave Function}
			& $\Psi_{\nu,k_y}(x,y;t) = e^{-i\epsilon t/\hbar}
			e^{ik_y y}e^{-\left(\theta/2\right)\sigma_y}\chi(x)$   
			& $\Psi_{\epsilon,k_{y}}(x,y;t) = e^{-i\epsilon t/\hbar}
			e^{ik_y y}e^{-\left(\theta/2\right)\sigma_y}e^{-i\left(\pi/4\right)\sigma_{y}}\chi(x)$ \\ \hline
			
			\parbox{2.5cm}{\raggedright Linearly Independent Solutions}
			& $\begin{aligned}
				\chi_{1}(x) &= 
				\begin{pmatrix}
					\operatorname{sgn}(\epsilon^\prime)\sqrt{\nu+1}D_{\nu}(\sqrt{2}\xi)\\[2pt]
					i D_{\nu+1}(\sqrt{2}\xi)
				\end{pmatrix} \\
				\chi_{2}(x) &= 
				\begin{pmatrix}
					-\operatorname{sgn}(\epsilon^\prime)\sqrt{\nu+1}D_{\nu}(-\sqrt{2}\xi)\\[2pt]
					i D_{\nu+1}(-\sqrt{2}\xi)
				\end{pmatrix}
			\end{aligned}$ 
			& $\begin{aligned}
				\chi_{1}(x) &= 
				\begin{pmatrix}
					e^{-i\xi^{2}/2}
					M\bigl(-il^{\prime2}_{E}k^{\prime2}_{y}/4,1/2;i\xi^{2}\bigr)\\[4pt]
					-l^{\prime}_{E}k^{\prime}_{y}\xi e^{-i\xi^{2}/2}
					M\bigl(1-il^{\prime2}_{E}k^{\prime2}_{y}/4,3/2;i\xi^{2}\bigr)
				\end{pmatrix} \\
				\chi_{2}(x) &= 
				\begin{pmatrix}
					-l^{\prime}_{E}k^{\prime}_{y}\xi e^{i\xi^{2}/2}
					M\bigl(1+il^{\prime2}_{E}k^{\prime2}_{y}/4,3/2;-i\xi^{2}\bigr)\\[4pt]
					e^{i\xi^{2}/2}
					M\bigl(il^{\prime2}_{E}k^{\prime2}_{y}/4,1/2;-i\xi^{2}\bigr)
				\end{pmatrix}
			\end{aligned}$ \\ \hline
			
			\parbox{2.5cm}{\raggedright Auxiliary Function}
			& $\xi(x)=x/l^{\prime}_{B}+l^{\prime}_{B}k^{\prime}_{y}$
			& $\xi(x)=x/l^{\prime}_{E}-\epsilon^{\prime}l^{\prime}_{E}/\hbar v_\text{F}$ \\ 
		\end{tabular}
	\end{ruledtabular}
\end{table*}

We start with the low-energy effective Hamiltonian
\begin{equation}
	H=v_\text{F}\boldsymbol{\sigma}
	\cdot
	\left(\boldsymbol{\hat{p}}+e\boldsymbol{A}\right)-e\varphi\sigma_{0},
	\label{eq:diracham}
\end{equation}
which describes massless relativistic Dirac fermions with Fermi velocity  $v_\text{F}$ under the in-plane electric field $\boldsymbol{E}=-\nabla\varphi=E\hat{x}$ and perpendicular magnetic field $\boldsymbol{B}=\nabla\times\boldsymbol{A}=B\hat{z}$. The scalar potential is chosen as $\varphi=-Ex$, and the vector potential is taken in the Landau gauge $\boldsymbol{A}=Bx\hat{y}$. 
In Eq.~(\ref{eq:diracham}), $\boldsymbol{\sigma}$ and $\sigma_{0}$ denote the Pauli and identity matrices, respectively, and the electron charge is $-e$.
To exploit the Lorentz covariance, we introduce the (2+1)-dimensional spacetime coordinate $x^{\mu=\left(0,1,2\right)}=\left(v_{\text{F}}t,x,y\right)$.
The manifestly covariant Dirac equation takes the following form
\begin{equation}
	\gamma^{\mu}\left(\hat{p}_{\mu}-eA_{\mu}\right)\Psi\left(x^{\mu}\right)=0,
	\label{eq:diraceq}
\end{equation} 
where the gamma matrices $\gamma^{\mu}=\left(\sigma_{z}, i\sigma_{y},-i\sigma_{x}\right)$ satisfying the Clifford algebra $\left\{ \gamma^{\mu},\gamma^{\nu}\right\} =2g^{\mu\nu}\sigma_{0}$ with the Minkowski metric $g^{\mu\nu}=\text{diag}\left(1,-1,-1\right)$, the spacetime momentum operator $\hat{p}_{\mu}=-i\hbar\partial_{x^{\mu}}$, and the electromagnetic  potential $A^{\mu}=\left({\varphi}/{v_\text{F}},A_{x},A_{y}\right)$. The metric $g^{\mu\nu}$ and its inverse $g_{\mu\nu}$ serve to raise and lower indices, thereby converting covariant vectors into contravariant ones and vice versa.
Repeated indices in Eq.~(\ref{eq:diraceq}) are summed over according to the Einstein summation convention.

It is instructive to show the general transformation of electromagnetic fields between different inertial frames.
The derivation is presented in Supplemental Materials~\cite{SM}.
Specifically, we consider the Lorentz boost with drift velocity $\boldsymbol{v}$ along the $\boldsymbol{E}\times\boldsymbol{B}$ direction, under which the contravariant coordinates transform as
\begin{equation}
	\left(\begin{array}{c}
		x^{\prime0}\\
		x^{\prime2}
	\end{array}\right)=\left(\begin{array}{ccc}
		\cosh\theta & \sinh\theta\\
		\sinh\theta & \cosh\theta
	\end{array}\right)\left(\begin{array}{c}
		x^{0}\\
		x^{2}
	\end{array}\right),
\end{equation}
and $x^{\prime1}=x^{1}$. 
Here, the drift velocity is parametrized as $v/v_\text{F}=\tanh\theta$, with the Fermi velocity $v_\text{F}$ playing the role of the universal speed limit, analogous to the speed of light in special relativity.
Conversely, the covariant momentum operator transforms under the inverse Lorentz transformation according to
\begin{equation}
	\left(\begin{array}{c}
		\hat{p}^{\prime}_{0}\\
		\hat{p}^{\prime}_{2}
	\end{array}\right)=\left(\begin{array}{cc}
		\cosh\theta & -\sinh\theta\\
		-\sinh\theta & \cosh\theta
	\end{array}\right)\left(\begin{array}{c}
		\hat{p}_{0}\\
		\hat{p}_{2}
	\end{array}\right),
\end{equation}
and $\hat{p}^{\prime}_{1}=\hat{p}_{1}$. 
This yields the relations
\begin{equation}
	\epsilon^{\prime}=\gamma\left(\epsilon+v\hbar k_{y}\right), \hbar k^{\prime}_{y}=\gamma\left(\hbar k_{y}+\frac{v}{v^{2}_{F}}\epsilon\right)
\end{equation}
with $\beta=\tanh\theta$ and $\gamma=1/\sqrt{1-\beta^{2}}$.
Accordingly, the spinor wave function in Eq.~(\ref{eq:diraceq}) transforms as
\begin{equation}
	\Psi^{\prime}\left(x^{\prime\mu}\right)
	=
	e^{\frac{\theta}{4}\left[\gamma^0,\gamma^2\right]}\Psi\left(x^{\mu}\right).
	\label{eq:lorentzwf}
\end{equation}
Consequently, the electromagnetic fields transform as \cite{Jackson2009}
\begin{equation}
	E^{\prime}=\gamma\left(E-vB\right), B^{\prime}=\gamma\left(B-\frac{v}{v^{2}_{F}}E\right).
	\label{eq:eb}
\end{equation}
For a prescribed electromagnetic field ratio, the drift velocity $v$, which cannot exceed the limiting Fermi velocity $v_\text{F}$, is chosen such that either the electric $\boldsymbol{E}^\prime$ or magnetic $\boldsymbol{B}^\prime$ field vanishes in Eq.~(\ref{eq:eb}). This dichotomy thus defines the magnetic and electric regimes, separated by the critical field ratio $E/B=v_\text{F}$.
The Dirac equation in Eq.~(\ref{eq:diraceq}) can be readily solved in the drifted inertial frames, where only either a magnetic field $\boldsymbol{B}^\prime$ or an electric field $\boldsymbol{E}^\prime$ remains~\cite{LANDAU1977,Andreev2007}, followed by an inverse Lorentz boost in Eq.~(\ref{eq:lorentzwf}) back to the laboratory frame. Without imposing boundary conditions, the general solutions, derived in Supplemental Materials~\cite{SM} and listed in Table~\ref{tab}, are a linear combination of two independent solutions.


\begin{figure}
	\centering
	\includegraphics[width=\columnwidth]{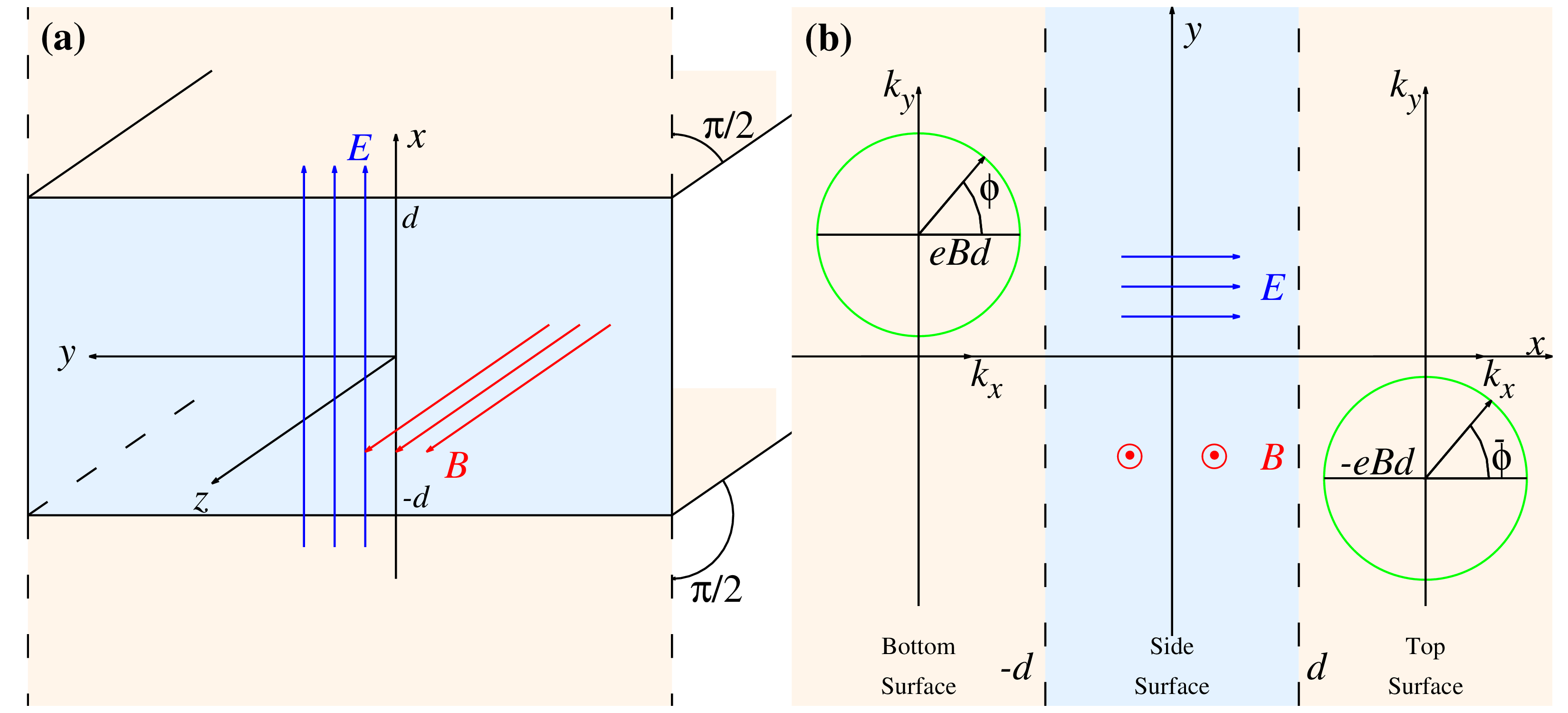}
	\caption{(a) Schematic of a 3D topological insulator slab under designed electromagnetic fields. (b) Effective unwrapped electromagnetic barriers with the in-plane electric $\boldsymbol{E}$ and perpendicular magnetic $\boldsymbol{B}$ field confined in the side surface region $\left|x\right|<d$.
	}
	\label{fig1}
\end{figure}

Having solved the Dirac equation, we are in a position to study the Klein tunneling through the electromagnetic barriers.
To this end, a three-dimensional topological insulator (TI) slab, as depicted in Fig.\ref{fig1}(a), serves as a natural setting with designed electromagnetic fields.
Three-dimensional TIs are bulk insulators with topologically protected conducting Dirac surface states~\cite{Hasan2010,Moore2010,Qi2011,Ando2013}.
In particular, strong TIs host an odd number of Dirac fermions on the surface, which are immune to perturbations~\cite{Bernevig2006,Bernevig2006a,Fu2007,Moore2007,Koenig2007,Hsieh2008,Roy2009}.
The low-energy surface Hamiltonian for a single Dirac fermion is given by
\begin{equation}
	H_{\text{surface}}
	=\frac{v_\text{F}}{2}\left(\boldsymbol{n}\cdot\left[\boldsymbol{\hat{p}}\times\boldsymbol{\sigma}\right]
	+\left[\boldsymbol{\hat{p}}\times\boldsymbol{\sigma}\right]\cdot\boldsymbol{n}\right),
\end{equation}
where $\boldsymbol{n}$ is the unit normal vector to the corresponding surface. 
To establish the connection with the effective Hamiltonian in Eq.~(\ref{eq:diracham}), two unitary transformations are introduced to unwrap the TI surfaces. First, the top and bottom surfaces are rotated by $\hat{R}_1\left(\theta\right)=\exp[-i\theta(\hat{S}_{y}+\hat{L}_{y})/\hbar]$ with $\theta=\pi/2$ and $-\pi/2$, respectively. Here, $\boldsymbol{\hat{S}}=\hbar\boldsymbol{\sigma}/2$ and $\boldsymbol{\hat{L}}=\boldsymbol{r}\times\boldsymbol{\hat{p}}$ denote the pseudospin and orbital angular momentum, respectively. As illustrated in Fig.~\ref{fig1}(a), all TI surfaces are mapped onto a single plane.
Second, a subsequent rotation $\hat{R}_2\left(\theta\right)=\exp[-i\theta\hat{S}_{z}/\hbar]$ with $\theta=\pi/2$ is performed in pseudospin space. 
The detailed derivation is presented in Supplemental Material~\cite{SM}.
As depicted in Fig.~\ref{fig1}(b), the resulting Hamiltonian recovers the form in Eq.~(\ref{eq:diracham}) with in-plane electric $\boldsymbol{E}$ and perpendicular magnetic $\boldsymbol{B}$ fields restricted to the side surface. 
The corresponding electromagnetic potentials are explicitly given by 
\begin{equation}
	\varphi=-E\times\begin{cases}
		-d & x<-d\\
		x & \left|x\right|<d\\
		d & x>d
	\end{cases},
	\boldsymbol{A}=B\hat{y}\times\begin{cases}
		-d & x<-d\\
		x & \left|x\right|<d\\
		d & x>d
	\end{cases}.
\end{equation}

Klein tunneling involves the process whereby a Dirac fermion is incident from the bottom surface, scattered by the electromagnetic potential in the side surface, and transmitted into the top surface. 
The derivation is presented in Supplemental Materials~\cite{SM}.
On both the bottom and top surfaces $\left|x\right|>d$, only free-propagating plane waves are retained, while evanescent waves are neglected because they do not affect the discussion of Klein tunneling.
On the side surface $\left|x\right|<d$, the wave function is constructed from the general solutions listed in Table~\ref{tab}, with the coefficients of linearly independent solutions determined by matching the wave functions at the interfaces $x=\pm d$. Below, we shall specify the details.
On the bottom surface ($x<-d$), an incident wave propagating to the right ($k^{\text{B}}_{x}>0$) with the energy dispersion  $\epsilon=v_{\text{F}}\sqrt{\left(\hbar k^{\text{B}}_{x}\right)^{2}+\left(\hbar k^{\text{B}}_{y}-eBd\right)^{2}}-eEd$ defines the incident angle $\phi$ via
\begin{equation}
	\tan\phi=\cfrac{\hbar k^{\text{B}}_{y}-eBd}{\hbar k^{\text{B}}_{x}}.
\end{equation}
The total wave function, including the incident and reflected ones, reads
\begin{equation}
	\chi_{\text{B}}(x)=\chi^{\text{in}}(x)+r\chi^{\text{re}}(x),
\end{equation}
where $r$ is the reflection coefficient and 
\begin{equation}
	\chi^{\text{in}}(x)=\cfrac{e^{ik^{\text{B}}_{x}x}}{\sqrt{2\cos\phi}}\left(\begin{array}{c}
		1\\
		e^{i\phi}
	\end{array}\right), \chi^{\text{re}}(x)=\cfrac{e^{-ik^{\text{B}}_{x}x}}{\sqrt{2\cos\phi}}\left(\begin{array}{c}
		1\\
		-e^{-i\phi}
	\end{array}\right).
\end{equation}
On the top surface ($x>d$), the transmitted wave with the energy dispersion 
\begin{equation}
	\epsilon=\pm v_{\text{F}}\sqrt{\left(\hbar k^{\text{T}}_{x}\right)^{2}+\left(\hbar k^{\text{T}}_{y}+eBd\right)^{2}}+eEd
\end{equation} 
defines the transmission angle $\bar{\phi}$
\begin{equation}
	\tan\bar{\phi}=\cfrac{\hbar k^{\text{T}}_{y}+eBd}{\hbar k^{\text{T}}_{x}},
\end{equation}
where the upper $+$ (lower $-$) sign corresponds to the electron (hole) state. 
This wave has the following form 
\begin{equation}
	\chi_{\text{T}}(x)=t\chi^{\text{tr}}(x),
\end{equation}
where $t$ is the transmission coefficient and 
\begin{equation}
	\chi^{\text{tr}}(x)=\cfrac{e^{ik^{\text{T}}_{x}x}}{\sqrt{2\left|\cos\bar{\phi}\right|}}\left(\begin{array}{c}
		1\\
		\pm e^{i\bar{\phi}}
	\end{array}\right),
\end{equation}
where $\cos\bar{\phi}>0$ ($<0$) for electron (hole) state.
The conservation of $y$-direction momentum, $k^{\text{B}}_{y}=k^{\text{T}}_{y}$, yields the angular relationship between the incident and transmitted waves
\begin{equation}
	\sin\bar{\phi}=\pm\cfrac{\epsilon+eEd}{\epsilon-eEd}\sin\phi\pm\cfrac{2v_{\text{F}}eBd}{\epsilon-eEd}.
	\label{eq:angle}
\end{equation}
It indicates that no transmission through the electromagnetic barrier is possible for certain incidence angles, generalizing the magnetic confinement discussed in Ref.~\cite{DeMartino2007}.
Detailed analysis reveals that the transmission into the hole state is strictly forbidden for any incident angle $\phi$ in the magnetic regime.
On the side surface $\left|x\right|<d$, the wave function can be written as
\begin{equation}
	\chi_{\text{S}}(x)=c_{1}\chi_{1}(x)+c_{2}\chi_{2}(x),
\end{equation}
where $\chi_{1,2}\left(x\right)$ are linearly independent solutions listed in Table~\ref{tab}. 
The continuity of wave functions at the interfaces of adjacent surfaces $x=\pm d$ yields the following relations 
\begin{equation}
	\begin{aligned}
		\left(\begin{array}{c}
			1\\
			r
		\end{array}\right)&=M\left(\begin{array}{c}
			c_{1}\\
			c_{2}
		\end{array}\right),
		\left(\begin{array}{c}
			c_{1}\\
			c_{2}
		\end{array}\right)&=Vt,
	\end{aligned}
	\label{eq:con}
\end{equation}
respectively.
Here, the transfer matrix is defined as
$M=\Xi^{-1}_{\text{B}}(-d)\Xi_{\text{S}}(-d)$
and the vector 
$V=\Xi^{-1}_{\text{S}}(+d)\chi^{\text{tr}}(+d)$
can be expressed in terms of the matrices composed of the spinor wave functions
$\Xi_{\text{B}}(x)=\left[\begin{array}{cc}
		\chi^{\text{in}}(x) & \chi^{\text{re}}(x)\end{array}\right]$
and 
$\Xi_{\text{S}}(x)=\left[\begin{array}{cc}
	\chi_{1}(x) & \chi_{2}(x)\end{array}\right]$.

\begin{figure}
	\centering
	\includegraphics[width=\columnwidth]{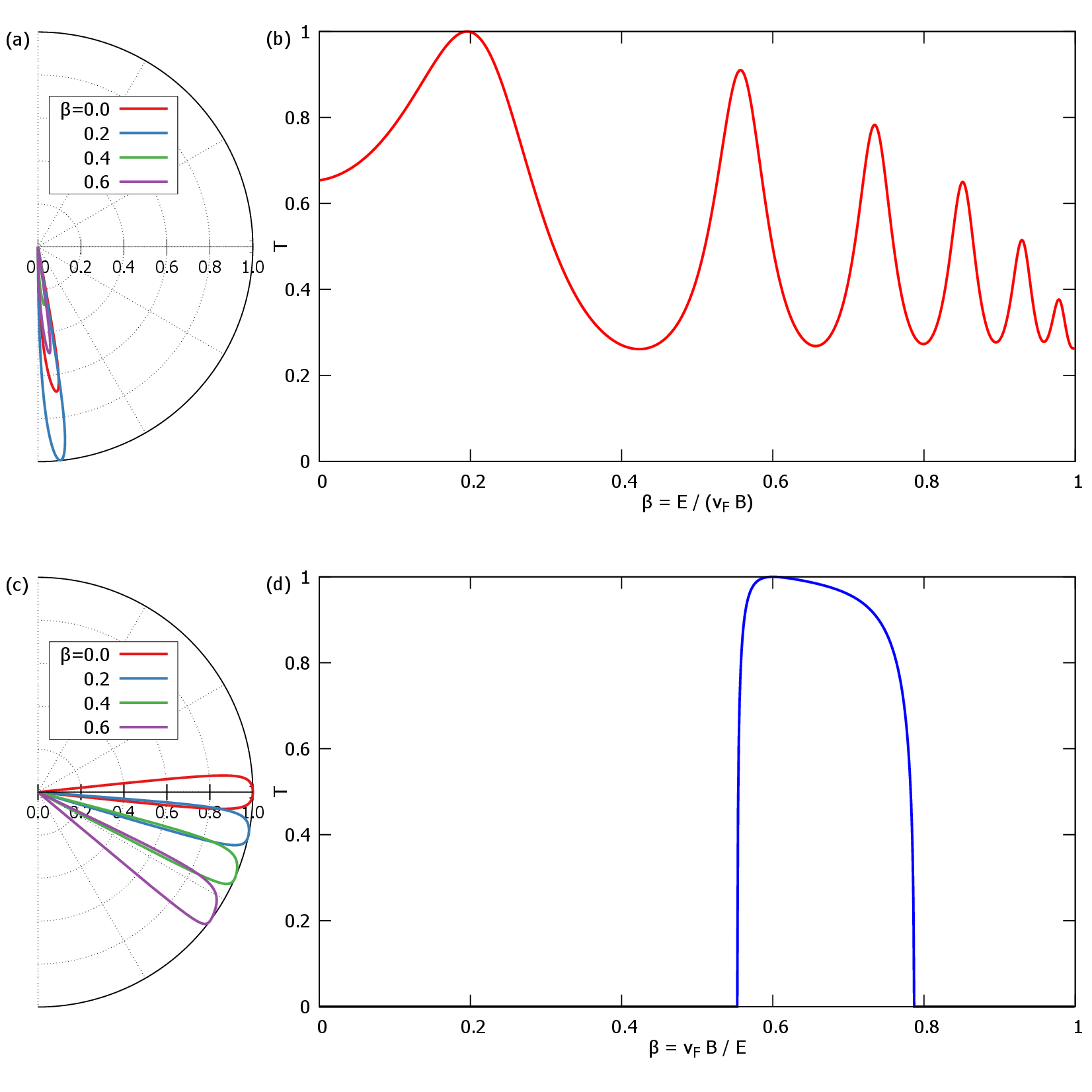}
	\caption{ 
		Polar graph of transmission probability $T(\phi)$ of Dirac fermions  through the electromagnetic barrier in the (a) magnetic and (c) electric regimes at fixed energy $\epsilon=3.7\hbar v_\text{F}/l_{B}$ and $\epsilon=3.7\hbar v_\text{F}/l_{E}$, and barrier width $d=3.67l_{B}$ and $d=3l_{E}$, respectively, for various $\beta$.
		The radial axis gives transmission $T$ ranging from $T=0$ at the center to $T=1$ at the outermost semicircle.
		Line plot of $T$ as a function of $\beta$ in the (b) magnetic and (d) electric regimes at fixed energy $\epsilon=3.7\hbar v_\text{F}/l_{B}$ and $\epsilon=3.7\hbar v_\text{F}/l_{E}$, incident angel $\phi=\arcsin(-0.995)$ and $\phi=\arcsin(-0.6)$, and barrier width $d=3.67l_{B}$ and $d=3l_{E}$, respectively.
		}
	\label{fig2}
\end{figure}

Ultimately, by solving the continuity equations in Eq.~(\ref{eq:con}), we arrived at the unified expression for the transmission probability
\begin{equation}
	\begin{aligned}
		T=\left|t\right|^{2}
		=	\cfrac{\left|\frac{1}{M_{11}}\right|^{2}\left|\frac{1}{V_{1}}\right|^{2}}{1+\left|\frac{M_{12}}{M_{11}}\right|^{2}\left|\frac{V_{2}}{V_{1}}\right|^{2}-2\left|\frac{M_{12}}{M_{11}}\right|\left|\frac{V_{2}}{V_{1}}\right|\cos\Phi}
	\end{aligned}
\end{equation}
with the Fabry-P\'erot interference phase
\begin{equation}
	\Phi=\arg\left(-\frac{M_{12}}{M_{11}}\right)+\arg\left(\frac{V_{2}}{V_{1}}\right),
	\label{eq:fp}
\end{equation}
irrespective of whether the transmitted wave is in the electron or hole states. Figures~\ref{fig2}(a) and~\ref{fig2}(c) show the calculated transmission $T$ versus incident angle $\phi$ for several sets of $\beta$, in the magnetic and electric regimes, respectively. 
In both regimes, quantum tunneling is forbidden over the vast majority of incident angles, as dictate by Eq.~(\ref{eq:angle}). 
In contrast, perfect Klein tunneling occurs only at specific $\beta$ in the magnetic regime, but is universal for general $\beta$ in the electric regime. This distinction is further confirmed by the numerical evaluation of transmission as a function of $\beta$ in Figs.~\ref{fig2}(b) and~\ref{fig2}(d), which reveals markedly different behaviors in these two regimes. The oscillatory behavior in the magnetic regime is expected from the Aharonov–Bohm effect in the drifted frame.
Strikingly, we obtain an analytical condition $\sin\phi=-\beta$ for the perfect Klein tunneling in the electric regime.
It corresponds to the normal incidence with $k^{\prime}_{y}=0$ in the drifted frame, generalizing the Klein collimation under the pure electric field~\cite{Cheianov2006}.

\begin{figure}
	\centering
	\includegraphics[width=\linewidth]{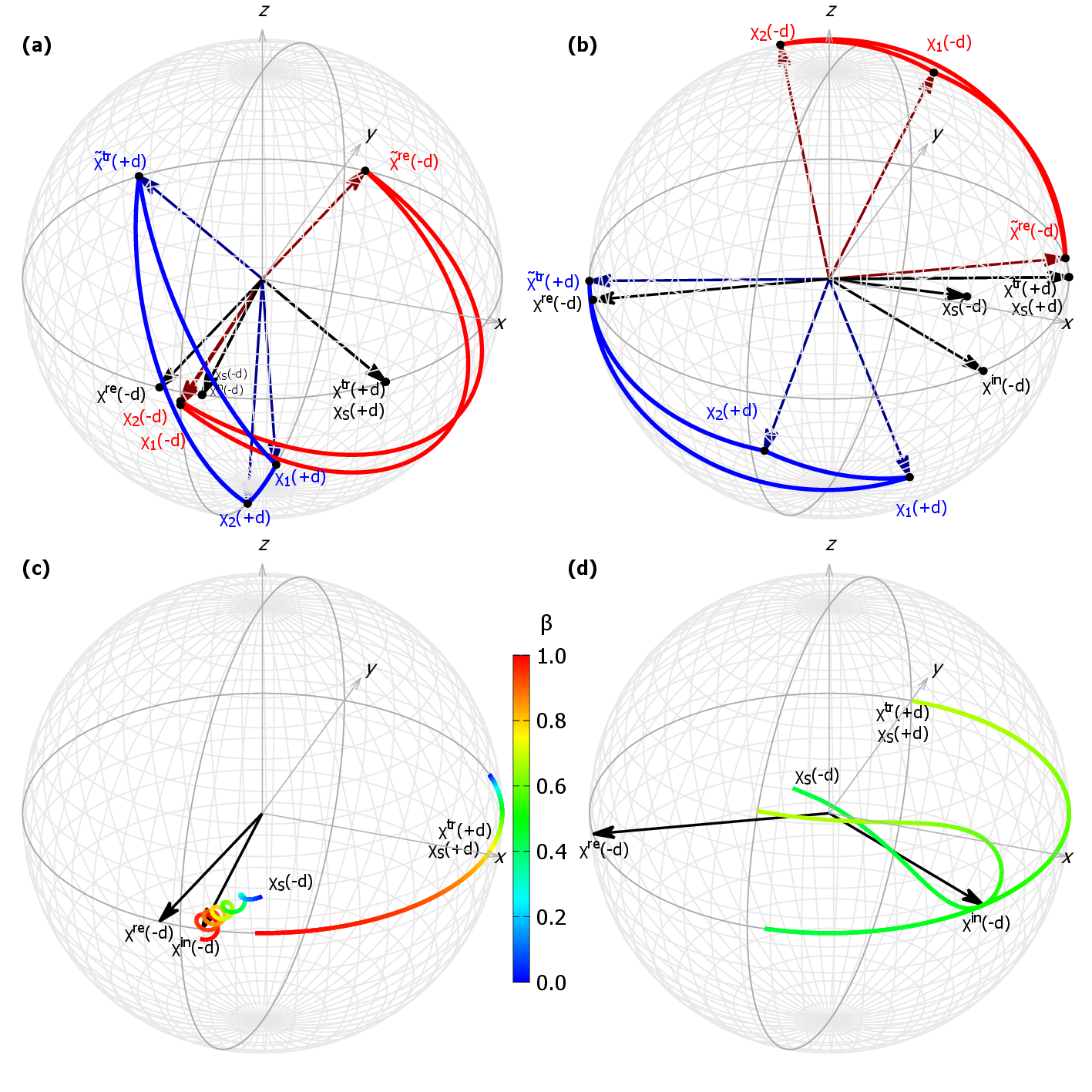}
	\caption{Geometric phase on the Bloch sphere. 
		The solid angles $\Omega\left(\pm d\right)$ covered by the geodesic triangles of spinors at the interfaces of adjacent surfaces $x=\pm d$ with fixed parameters $\{\phi, \epsilon l_{B}/\hbar v_{\text{F}}, d/l_{B}, \beta\}=\{\arcsin(-0.995), 3.7, 3, 0.953\}$
		and $\{\phi, \epsilon l_{E}/\hbar v_{\text{F}}, d/l_{E}, \beta\}=\{\arcsin(-0.5), 3.7, 3, 0.6\}$ in the (a) magnetic and (b) electric regimes, respectively. The corresponding trajectories of pseudospin polarization in the (c) magnetic and (d) electric regimes with $\beta$ released.}
	\label{fig3}
\end{figure}

To gain further insights, the Fabry–P\'erot interference phase in Eq.~(\ref{eq:fp}) can be reexpressed as
\begin{eqnarray}
	\Phi=
	&-&\arg\langle\chi_{2}(-d)\vert\chi_{1}(-d)\rangle
	-\cfrac{\Omega\left(-d\right)}{2} \nonumber \\
	&+&\arg\langle\chi_{2}(+d)\vert\chi_{1}(+d)\rangle
	+\cfrac{\Omega\left(+d\right)}{2}
\end{eqnarray}
in terms of spinor wave functions at the interfaces of adjacent surfaces $x=\pm d$~\cite{Choi2013}. 
Here the solid angle $\Omega\left(\pm d\right)$ are covered by the geodesic triangle connecting the spinors $\chi_{1}(\pm d)$, $\chi_{2}(\pm d)$, and $\tilde{\chi}^{\text{re/tr}}(\pm d)$ on the Bloch sphere, where $\tilde{\chi}^{\text{re/tr}}$ are the spinors orthogonal to $\chi^{\text{re/tr}}$. It can be shown that the inner products $\langle\chi_{2}(\pm d)\vert\chi_{1}(\pm d)\rangle$ are purely real and hence have no contributions.
To geometrically visualize the phase, the relevant spinor $\chi$ can be mapped onto the pseudospin polarization vector $\boldsymbol{P}={\langle\chi\vert\boldsymbol{\sigma}\vert\chi\rangle}/{\langle\chi\vert\chi\rangle}$ on the Bloch sphere.
Figures~\ref{fig3}(a) and \ref{fig3}(b) display the geodesic triangles that define the solid angles $\Omega\left(\pm d\right)$ in the magnetic and electric regimes, respectively.  
Notably, under the perfect transmission condition $\sin\phi=-\beta$ in the electric regime, three vertices become coplanar and the solid angle collapses to zero, consistent with the vanishing of reflection-induced phase shift.
To examine how the solid angles evolve as $\beta$ varies, we trace the polarization trajectories on the Bloch sphere, as sketched in Figs.~\ref{fig3}(c) and \ref{fig3}(d), for the magnetic and electric regimes, respectively.
The transmitted trajectory intersects the incident polarization at perfect transmission, which implies that the chirality of Dirac fermions is conserved in Klein tunneling.

Before closing, we briefly discuss the titling of Dirac cones, described by the following Hamiltonian
\begin{equation}
	H=v_\text{F}\boldsymbol{\sigma}\cdot
	\left(\boldsymbol{\hat{p}}+e\boldsymbol{A}\right)
	+\boldsymbol{w}\cdot\boldsymbol{\hat{p}}\sigma_{0},
\end{equation}
where the tilting vector is chosen along the $y$-axis with the form $\boldsymbol{w}=w_{y}\hat{y}$. 
Under the Landau gauge $\boldsymbol{A}=Bx\hat{y}$, $p_{y}$ is a good quantum number. Shifting the spectrum, we arrived at the effective Hamiltonian
\begin{equation}
	H-\omega_{y}p_{y}\sigma_{0}
	=v_\text{F}\left[\hat{p}_{x}\sigma_{x}+\left(p_{y}+eBx\right)\sigma_{y}\right]+\omega_{y}eBx\sigma_{0},
\end{equation}
which describes the Dirac fermions under an effective electric field $\boldsymbol{E}=\omega_{y}B\hat{x}$ and a magnetic field $\boldsymbol{B}=B\hat{z}$ in Eq.~(\ref{eq:diracham}).
Therefore, a relation between the tilting and the electric field can be established.
The Landau levels stabilize in the undertilted ($w_{y}/v_\text{F}<1$) regime and collapse in the overtilted ($w_{y}/v_\text{F}>1$) regimes~\cite{Yu2016,Udagawa2016,Tchoumakov2016}, corresponding to the magnetic and electric regimes discussed above, respectively.

To summarize, the Dirac equation under both the in-plane electric $\boldsymbol{E}$ and perpendicular magnetic $\boldsymbol{B}$ fields is solved by taking advantage of Lorentz covariance. 
Then, we present a comprehensive study of Klein tunneling through generalized electromagnetic potentials, in which the Fabry-P\'erot interference is further analyzed using the geometric representation of spinor wave functions on the Bloch sphere. Finally, we briefly discuss the tilting of Dirac cones in relation to the in-plane electric field.
Our study explores Klein tunneling across a generalized heterojunction with generic electromagnetic potentials, yielding insights into both the fundamental physics and the design of novel electronic devices.

This work is supported by National Natural Science Foundation of China under Grants No. 12174345, and Zhejiang Provincial Natural Science Foundation of China under Grant No. LZ22A040002.

	
	%

\end{document}